\documentclass[twocolumn, superscriptaddress, aps, prx]{revtex4-2}
\usepackage{amsmath,amssymb, amsfonts}
\usepackage{graphicx}
\usepackage{pgfplots}
\usepackage{algorithm}
\pgfplotsset{compat=1.18}
\usepackage{tikz}
\usepackage{bm}
\usepackage{hyperref}
\usepackage{xcolor}
\usepackage{newtxtext,newtxmath}
\usepackage{algorithmicx}
\usepackage{algpseudocode}
\usepackage{textcomp}
\usepackage{float}
\usepackage{booktabs}
\usepackage{microtype}
\usepackage{orcidlink}
\usepackage{comment}
\usepackage[normalem]{ulem}

\hypersetup{
    colorlinks=true,
    allcolors=blue,
    urlcolor=blue,
    citecolor=blue,
    linkcolor=blue
}

\makeatletter
\AtBeginDocument{%
  \renewcommand{\@makecaption}[2]{%
    \vspace{\abovecaptionskip}%
    \leftskip=0.5em\rightskip=0.5em\parfillskip=0pt plus 1fil%
    {\small\textbf{#1.}\ #2\par}%
    \vspace{\belowcaptionskip}%
  }%
}
\makeatother

\pgfplotsset{every axis/.append style={line width=1pt, tick style={line width=0.8pt}}}

\newcommand{\comments}[1]{}

\begin{document}

\title{Hybrid Quantum-Classical Machine Learning Algorithms \\for Multi-Output Time-Series Forecasting at Utility Scale}

\author{{Mackenson Polché}\orcidlink{0009-0008-1789-0712}}
\email{mackenson.p@thewiser.org}
\affiliation{The Washington Institute for STEM, Entrepreneurship, and Research (WISER), Washington DC, USA}
\affiliation{Department of Basic and Applied Sciences, CUTonalá, Universidad de Guadalajara, Av. Nuevo Periférico No. 555, Ejido San José Tateposco, Tonalá 45425, Mexico}
\author{{Varun Puram}\orcidlink{0000-0002-3729-671X}}
\affiliation{The Washington Institute for STEM, Entrepreneurship, and Research (WISER), Washington DC, USA}
\affiliation{Oklahoma State University, Department of Computer Science, Stillwater, USA}
\author{{Aditi Lal}\orcidlink{0009-0006-7985-2098}}\affiliation{The Washington Institute for STEM, Entrepreneurship, and Research (WISER), Washington DC, USA} \affiliation{Department of Applied Mathematics, South Asian University, New Delhi, India}
\author{{Weronika Golletz}\orcidlink{0000-0002-8639-0227}}
\author{Joan Étude Arrow}
\author{{Vardaan Sahgal}\orcidlink{0000-0002-2293-7430}}
\email{vardaan.s@thewiser.org}
\affiliation{The Washington Institute for STEM, Entrepreneurship, and Research (WISER), Washington DC, USA}
\author{{Kumar Ghosh}\orcidlink{0000-0002-4628-6951}}
\author{{Giorgio Cortiana}\orcidlink{0000-0001-8745-5021}}
\author{{Corey O'Meara}\orcidlink{0000-0001-7056-7545}}
\affiliation{E.ON Digital Technology GmbH, Hannover, Germany}

\begin{abstract}
Multi-output time-series forecasting in energy systems is challenging because of nonlinear dynamics, multi-scale seasonality, and strong dependencies across correlated series. In this work, we investigate two hybrid quantum-classical frameworks for multi-stream time-series forecasting on a real Smart Meter dataset comprising 103 household electricity consumption time-series, with experiments executed on the \texttt{ibm\_marrakesh} superconducting quantum processor.  The first model, Kernelized Quantum Reservoir Computing with Repeated Measurement (KQRC-RM), combines coupled quantum reservoirs, ancilla-assisted repeated measurement, and kernelized readouts to model temporal dynamics and cross-stream correlations jointly. For a 3-stream time-series input and output, the KQRC-RM model using 114 qubits achieves an MAE of 0.0811 on MPS simulator (36.92\% improvement over its classical analog) whereas performance degrades to an MAE of 0.1524 on hardware. The second, a Projected Quantum Kernel Gaussian Process (QGP), replaces fidelity-based kernels with projected kernels constructed from local reduced-state statistics. Using a topology-aware 100-qubit QGP model to predict 100 multi-output time-series values, we observe 49\% of time-series outputs achieve high-accuracy predictions (MAE $<0.15$), with an average MAE of $0.082$ for this low-error group. The medium-error regime (MAE $0.15$–$0.35$) has an average MAE of $0.229$, while the high-error regime (MAE $>0.35$) has an average MAE of $0.664$. Overall, this reduces the average MAE relative to the classical GP baseline by 62.01\% on MPS simulator and 40.37\% on hardware. Together, these results demonstrate the feasibility of hybrid quantum machine learning for multi-input, multi-output time-series forecasting at the 100+ qubit scale on NISQ devices. 

\end{abstract}

\maketitle

\section{Introduction}
\label{sec:introduction}

Accurate time-series forecasting underpins the reliable operation of modern energy systems, including load balancing, unit commitment, and the integration of variable renewable resources.
In practice, forecasting electricity demand and related energy indicators remains challenging because these signals exhibit strong temporal dependencies, multi-scale seasonality, non-stationarity, and variability driven by consumer behavior and weather conditions~\cite{Weron2006}.
Classical statistical models such as ARIMA and state-space methods rely on linear or weakly nonlinear assumptions and often struggle in renewable-heavy smart-grid settings~\cite{Boxetal}.
More flexible machine-learning approaches can improve predictive performance, but they typically require substantial training data and computational resources, particularly for high-dimensional multivariate forecasting problems~\cite{hong}.

Quantum machine learning offers an alternative set of modeling tools for such problems.
The motivation for studying quantum methods in forecasting is not that quantum advantage is already established in this domain, but rather that quantum feature maps, reservoir dynamics, and variational circuits can induce expressive nonlinear representations of data and similarity measures that may be difficult to construct classically ~\cite{ahmed2024,juarez2025}.
At the same time, practical quantum learning on current Noisy Intermediate-Scale Quantum (NISQ) devices remains constrained by limited qubit counts, shallow circuit depths, and hardware noise.
As a result, near-term applications are most realistically pursued in hybrid quantum-classical settings, where quantum circuits act as feature-generating or nonlinear processing components within otherwise classical learning pipelines.

Within this hybrid setting, two quantum learning paradigms are particularly relevant for structured temporal forecasting.
The first is Quantum Reservoir Computing (QRC), introduced by Fujii and Nakajima~\cite{b35}, in which quantum dynamics are used as a fixed nonlinear reservoir for time-series processing.
Later work extended QRC across different physical platforms~\cite{b36,b37}, and Yasuda et al.~\cite{b38} showed that repeated measurement can introduce useful nonlinear transformations for temporal prediction on superconducting hardware.
However, existing QRC approaches remain largely focused on single-stream settings and typically use simple classical readouts.

The second relevant paradigm is quantum kernel learning.
Quantum feature maps can be interpreted as implicit kernels in Hilbert space~\cite{b39}, and kernel methods based on circuit-induced similarities have shown promise for supervised learning~\cite{b40}.
However, fidelity-based quantum kernels require deep compute--uncompute circuits, are sensitive to hardware noise, and may suffer from exponential concentration with increasing system size~\cite{b16,b17}.
Projected quantum kernels alleviate these limitations by replacing global state overlaps with statistics of local observables derived from reduced density matrices, thereby improving robustness on near-term devices~\cite{b16}.
Although projected kernels have been studied for classification and other regression tasks~\cite{b18,b19,b20}, their use in structured multi-output time-series forecasting remains largely unexplored.

In this work, we investigate these two directions through two complementary quantum frameworks for multi-stream energy time-series forecasting.
The first is a Kernelized Quantum Reservoir Computing with Repeated Measurement (KQRC-RM) architecture~\cite{qrc}, which combines coupled quantum reservoirs, repeated measurement feedback, and kernelized readouts to model both temporal dynamics and cross-stream correlations.
Related recent work by Dao et al.~\cite{dao2026breaking} considers a single-stream quantum extreme learning machine based on a kicked-Ising reservoir with a classical linear readout.
However, that setting is limited to single-channel benchmarks and does not address kernelized multi-stream forecasting of correlated energy-demand series.
To the best of our knowledge, prior work has not unified quantum reservoir dynamics~\cite{b35,b36,b37}, repeated-measurement feedback~\cite{b38}, and quantum kernel regression~\cite{b39,b40} in a single multi-stream forecasting architecture.

The second framework is a Projected Quantum Kernel Gaussian Process (QGP), which replaces fidelity-based kernels with kernels constructed from local expectation values, enabling a more noise-resilient Bayesian approach to multi-output regression.

Both frameworks are evaluated on real quantum hardware and compared against established classical baselines.
In this way, the paper provides an empirical study of how kernelized and reservoir-based quantum models behave on practical energy forecasting tasks under realistic NISQ constraints.

The paper is organized as follows.
In Section~\ref{sec:method}, we introduce the Smart Meter dataset, the KQRC-RM and QGP models, and the theoretical framework used to assess quantum performance.
Section~\ref{sec:results} presents the main experimental results for both models on simulator and real quantum hardware, including the large-scale QGP experiment.
In Section~\ref{sec:benchmarking}, we compare the proposed quantum approaches with classical forecasting baselines.
Finally, Section~\ref{sec:conclusion} concludes the paper.
Supporting methodological details and supplementary results are provided in the Appendix.

\section{Methods}
\label{sec:method}
In this section, we describe the dataset, the two quantum forecasting models used in this work, KQRC-RM and QGP, and the theoretical framework used to assess quantum performence.

\subsection{Dataset}
\label{sec:dataset}

We employ an anonymized Smart Meter dataset comprising hourly electricity-consumption
measurements from 103 residential customers during the winter season. Each observation
records the energy consumed at a given timestamp, yielding a collection of univariate time-series. A representative sample is shown in
Appendix~\ref{app:dataset} (Fig.~\ref{fig:splitting_customers}). 

To enable joint modelling, customer subsets are constructed according to pairwise correlation strength. For the QGP experiments, we select a subset of 15 highly correlated customers from the full dataset. For the KQRC-RM experiments, smaller subsets are used because of the increased readout complexity, i.e., triplets of customers are sampled from the highly correlated group.
Further details on the subset construction and data splitting procedure are provided in Appendix~\ref{app:dataset}.

Throughout this work, the multivariate observation at time $t$ is denoted by
\[
x_t = \bigl(x_t^{(1)}, x_t^{(2)}, \ldots, x_t^{(S)}\bigr),
\]
where $S$ is the number of jointly modelled customers. The forecasting task is to predict future electricity demand by exploiting both the temporal structure within each series and the correlations across series.

\subsection{Kernelized Quantum Reservoir Computing with Repeated Measurement}
\label{sec:kqrc}

\begin{figure*}[!bth]
\centering
\includegraphics[]{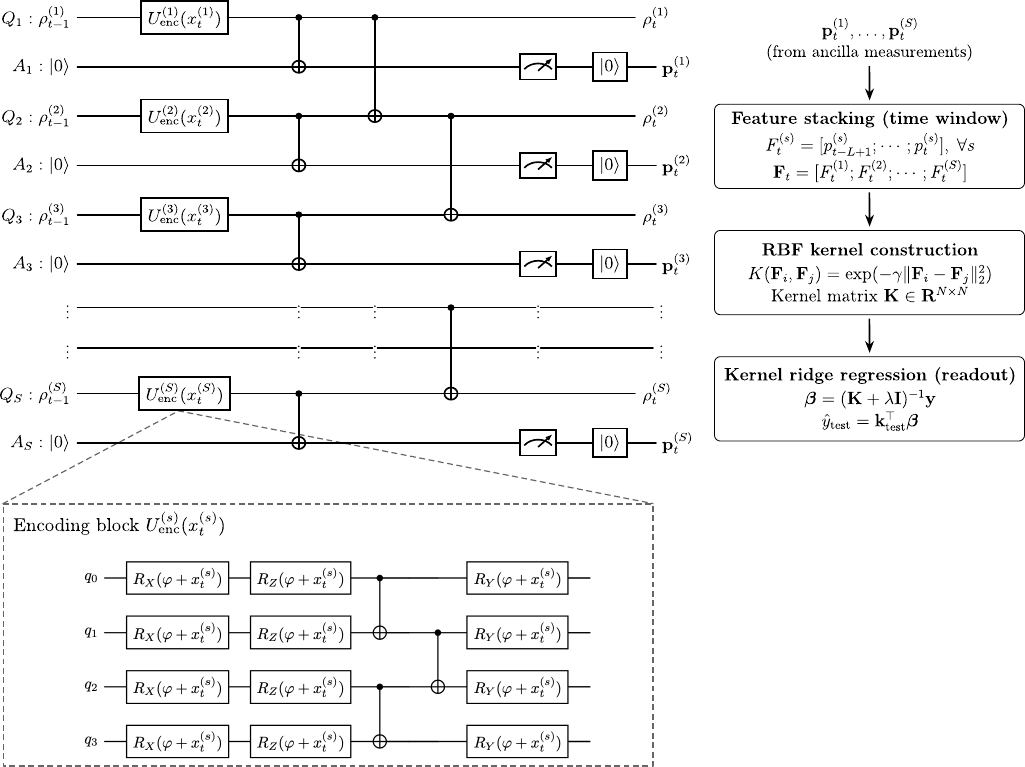}
\caption{
Circuit-level view of the proposed Kernelized Quantum Reservoir with repeated
measurement. Each input stream $x_t^{(s)}$ is encoded into its system register
($q_{s,0},\ldots,q_{s,n_q-1}$) via a parametrized encoding unitary $U_{\mathrm{enc}}^{(s)}$. Intra-stream entangling gates propagate information within each register, while inter-stream couplings between boundary qubits introduce cross-stream interactions. 
Each system qubit is then coupled to a corresponding ancilla qubit
($a_{s,0},\ldots,a_{s,n_q-1}$), and only the ancillas are measured at each timestep.
Repeated measurements produce probability vectors $\mathbf{p}_t^{(s)}$ that are accumulated into a per-stream feature matrix and used to construct an RBF
kernel $K^{(s)}$ for kernel ridge regression. The inset shows the example for $n_q=4$ qubits of the parametrized encoding block $U_{\mathrm{enc}}^{(s)}$ used within each stream.}
\label{fig:arch}
\end{figure*}

Quantum reservoir computing~\cite{b35,b38,b39,b40} exploits the dynamics of quantum systems as high-dimensional feature maps for time-series processing, while only a classical readout layer is trained. In this framework, input data are sequentially encoded into a quantum system, whose evolving state captures temporal information. Here, we extend this paradigm to multiple correlated time-series and combine it with repeated ancilla-based measurements and kernel ridge regression.

\subsubsection{Reservoir Architecture and Encoding}

We consider $S$ correlated time-series
$\{x_t^{(1)}, x_t^{(2)}, \ldots, x_t^{(S)}\}$
sampled at discrete timesteps $t = 1, 2, \ldots, T$.
In our setting, each series corresponds to the energy consumption of a distinct customer, although the framework readily extends to other types of correlated signals (e.g., voltage stability or generation output).

For each time-series $s$, we allocate a register of $n_q$ system qubits
$\{ q_{s,0}, q_{s,1}, \ldots, q_{s,n_q-1} \}$
and a corresponding register of $n_q$ ancilla qubits
$\{ a_{s,0}, a_{s,1}, \ldots, a_{s,n_q-1} \}$.
Each logical channel consists of a system register and
a matched ancilla register used for readout. The circuit-level architecture is illustrated in
Fig.~\ref{fig:arch}, where each system line ($Q$\, corresponding to qubits $q_{s,i}$) is paired with an ancilla line ($A$, corresponding to qubits $a_{s,i}$), The system qubits store and evolve the quantum state (quantum memory), while the ancilla qubits are used for measurement and readout, and only the ancilla qubits are measured at each timestep.

At timestep $t$, the observation vector $x_t$ is embedded into the quantum state through a parameterized encoding
unitary. For each series $s$, a dedicated circuit
$U_{\mathrm{enc}}^{(s)}\!\left(x_t^{(s)}\right)$ acts on the corresponding system register, combining
data-dependent single-qubit rotations with a fixed brickwork entanglement structure. The
full encoding unitary is given by:
\begin{equation}
U_{\mathrm{enc}}(x_t)
=
\bigotimes_{s=1}^{S}
U_{\mathrm{enc}}^{(s)}\!\left(x_t^{(s)}\right),
\label{eq:uenc}
\end{equation}
where each $U_{\mathrm{enc}}^{(s)}$ consists of one layer of $R_X$, $R_Z$, and
$R_Y$ rotations interleaved with a brickwork CNOT block, as illustrated in
Figure~\ref{fig:arch} for $n_q=4$ qubits. 

Let $\rho_{t-1}$ denote the post-measurement system state carried over from timestep $t-1$, with $\rho_0$ initialized in the computational zero state. Applying the encoding unitary yields:
\begin{equation}
\rho_t'
=
U_{\mathrm{enc}}(x_t)\,
\rho_{t-1}\,
U_{\mathrm{enc}}^{\dagger}(x_t).
\end{equation}
The encoded state is then processed through entangling dynamics that couple information both within and across time-series.

\subsubsection{Cross-Series Entangling Dynamics}

After encoding, entangling interactions are introduced to model dependencies among the time-series. As illustrated in Fig.~\ref{fig:arch}, these interactions act (i) within each individual series, and (ii) across neighboring series.

Let $q_{s,n_q-1}$ denote the tail qubit of series $s$ and $q_{s+1,0}$ the
head qubit of the next series $s{+}1$. Two-qubit entangling gates are applied along these links. The resulting reservoir unitary is 
\begin{equation}
U_{\mathrm{res}}
=
\prod_{s=1}^{S-1}
U_{\mathrm{inter}}\!\left(q_{s,n_q-1},\, q_{s+1,0}\right)
\cdot
\prod_{s=1}^{S} U_{\mathrm{intra}}^{(s)},
\label{eq:ures}
\end{equation}
where $U_{\mathrm{intra}}^{(s)}$ describes intra-series entanglement within series~$s$, and
$U_{\mathrm{inter}}$ represents the inter-series entangling operation acting on the neighbouring boundary qubits $q_{s,n_q-1}$ and $q_{s+1,0}$.

The system state after reservoir evolution is
\begin{equation}
\rho_t^{\rm res}
=
U_{\mathrm{res}}\,
\rho_t'\,
U_{\mathrm{res}}^{\dagger}.
\label{eq:rho_after_res}
\end{equation}
Equations~(\ref{eq:ures}) and~(\ref{eq:rho_after_res}) define a
multi-time-series quantum reservoir in which information  propagates both across time and across correlated input channels.

\subsubsection{Ancilla-Based Readout and Kernelized Regression}

Following the reservoir evolution, each system qubit $q_{s,i}$ is coupled to its
paired ancilla $a_{s,i}$ through a readout unitary  $U_{\mathrm{SA}}$. 
The joint system--ancilla state before measurement is
\begin{equation}
\tilde{\rho}_t
=
U_{\mathrm{SA}}
\Bigl(
\rho_t^{\mathrm{res}}
\otimes
|0\cdots0\rangle\!\langle 0\cdots0|_{\mathrm{anc}}
\Bigr)
U_{\mathrm{SA}}^\dagger.
\end{equation}
Only the ancilla qubits
are measured, while the system qubits retain coherence to carry memory across
timesteps.

Let $m_t$ denote the full ancilla measurement outcome at time~$t$. 
The corresponding measurement operator, extended from the single-series formulation 
of Yasuda et al.~\cite{b38}, is
\begin{equation}
M_{m_t}
=
I_{\text{sys}}
\otimes
\bigotimes_{s=1}^{S}
\bigotimes_{i=0}^{n_q-1}
\big|m_{t,s,i}\big\rangle
\!\big\langle m_{t,s,i}\big|,
\end{equation}
where $m_{t,s,i} \in \{0,1\}$ is the outcome obtained on ancilla $a_{s,i}$.
The conditional post-measurement state of the system is 
\begin{equation}
\rho_t^{(m_t)}
=
\frac{{\text{Tr}_{\rm anc} [}
M_{m_t}\,
\tilde{\rho}_t\,
M_{m_t}^{\dagger}]
}{
p(m_t)
},
\quad
p(m_t)
=
\mathrm{Tr}\!\left[
M_{m_t}\,
\tilde{\rho}_t
\right].
\label{eq:conditional_state}
\end{equation}
 The system state passed to the next timestep is the ensemble-average post-measurement state
\begin{equation}
\rho_t
=
\sum_{m_t} p(m_t)\,\rho_t^{(m_t)}.
\end{equation}

Repeated circuit execution (shots) yield an empirical probability
distribution over ancilla bitstrings,
\begin{equation}
\mathbf{p}_t
=
\left[
p_t(b_1),
p_t(b_2),
\ldots,
p_t(b_{N_b})
\right]^{T},
\label{eq:pt}
\end{equation}
where each $b_j$ denotes a bitstring outcome and $N_b$ is the number of tracked basis
states. After each measurement, the ancillas are reset to $\lvert 0\dots 0\rangle$, while the system state continues recurrently through the update above. 

For downstream regression, we form stream-specific probability vectors $\mathbf{p}_t^{(s)}$ by marginalizing the joint ancilla distribution $\mathbf{p}_t$ over all ancilla qubits not belonging to stream $s$. These distributions define the feature matrices
\begin{equation}
\mathbf{F}^{(s)} =
\begin{bmatrix}
\mathbf{p}_1^{(s)} \\
\mathbf{p}_2^{(s)} \\
\vdots \\
\mathbf{p}_T^{(s)}
\end{bmatrix},
\quad s = 1, 2, \ldots, S.
\label{eq:feature_mats}
\end{equation}

For each time-series $s$, we then construct a radial basis function (RBF) kernel:
\begin{equation}
K^{(s)}(i,j)
=
\exp
\!\Big(
-\gamma
\,
\Vert
\mathbf{p}_i^{(s)} - \mathbf{p}_j^{(s)}
\Vert_2^2
\Big),
\label{eq:rbf_kernel}
\end{equation}
where $\gamma>0$ is the bandwidth parameter. 
A ridge regression model is trained
in this feature space. Let $\mathbf{y}_{\text{train}}^{(s)}$ denote the training targets for series $s$. The  regression coefficients are 
\begin{equation}
\boldsymbol{\beta}^{(s)}
=
\big(
K_{\text{train}}^{(s)}
+
\lambda I
\big)^{-1}
\mathbf{y}_{\text{train}}^{(s)},
\label{eq:beta}
\end{equation}
where $\lambda$ is the regularization parameter. Predictions for the test set are then given by
\begin{equation}
\hat{\mathbf{y}}_{\text{test}}^{(s)}
=
K_{\text{test}}^{(s)}
\,
\boldsymbol{\beta}^{(s)}.
\label{eq:prediction}
\end{equation}

This framework yields a Kernelized Quantum Reservoir with repeated measurement that captures temporal structure through iterative evolution of the same system
qubits, and cross-stream correlations through entanglement between
neighboring streams.
Figure~\ref{fig:arch} summarizes the full workflow.

\subsection{Gaussian Process with a Projected Quantum Kernel}
\label{sec:qgp}

As a complementary Bayesian alternative to the recurrent KQRC-RM framework, we consider a Gaussian Process equipped with a projected quantum kernel. Whereas KQRC-RM learns a classical readout from recurrent reservoir states, the QGP model measures similarity directly between encoded inputs through local reduced-state statistics and naturally provides predictive uncertainty.

Gaussian Processes~\cite{2006Rasmussen,
2007Bonilla, 2012Alvarez} are flexible nonparametric models for regression, but their performance depends critically on the choice of kernel
function~\cite{b1,b3}. Quantum kernels derived from parametrized quantum feature maps can capture structure that  may be difficult for classical kernels to
represent~\cite{b6}. A common choice is the fidelity kernel, which quantifies similarity through the overlap of full quantum states.
However, fidelity-based kernels require deep compute--uncompute
circuits, are sensitive to noise on near-term quantum devices, and may suffer from
measurement-induced exponential concentration~\cite{b16,b17}. 
To address these limitations, we adopt a projected quantum kernel based on local reduced density matrices. This construction
compares multivariate inputs through local reduced states, can be estimated directly from measurement outcomes, and is used within a shared-kernel GP framework to predict multiple output series~\cite{b16,b17}.

\subsubsection{Quantum Gaussian Process}

For a multi-input, multi-output setting, let 
$X = \{\mathbf{x}_i\}_{i=1}^{N}$ denote the training inputs, where each $\mathbf{x}_i$ is a multivariate input vector, and 
$Y = [\mathbf{y}^{(1)},\ldots,\mathbf{y}^{(D)}] \in \mathbb{R}^{N\times D}$ 
denote the corresponding multi-output training targets, where $D$ is the number of outputs. 
The test inputs are denoted by $X_*$.

We adopt an independent-output Gaussian Process formulation, in which all outputs are predicted using a shared projected kernel while conditional independence across outputs is assumed given the inputs.
The GP uses a projected quantum kernel, denoted by 
$K_{\mathrm{proj}}^{(k)}(x_i,x_j)$ and defined in the next subsection, 
to construct the Gram matrix $K_{XX}$.

The predictive distribution of the GP at the test inputs is Gaussian, 
with mean and covariance given by
\begin{align}
\boldsymbol{\mu}_* 
&= 
K_{X_* X}
\bigl(K_{XX} + \sigma^{2} I\bigr)^{-1}
Y,
\\
\Sigma_* 
&= 
K_{X_* X_*} 
- 
K_{X_* X}
\bigl(K_{XX} + \sigma^{2} I\bigr)^{-1}
K_{XX_*},
\end{align}
where $\sigma^2$ is the observation-noise variance. 
The predictive mean $\boldsymbol{\mu}_* \in \mathbb{R}^{N_* \times D}$ 
provides simultaneous predictions for all outputs. The matrix $\Sigma_*$ is the predictive covariance in input space and is shared across outputs under the independent-output assumption. 
For each output dimension $d=1,\ldots,D$, the predictive variance is 
obtained from the diagonal elements of $\Sigma_*$.

The kernel parameters, together with the noise variance, are 
optimized by maximizing the joint GP marginal log-likelihood 
across all outputs~\cite{b12}:
\begin{equation}
\begin{aligned}
\log p(&Y\mid X,\theta)
=
-\frac{1}{2}
\sum_{d=1}^{D}
\mathbf{y}^{(d)\top}
\big(K_{XX}(\theta)+\sigma^{2}I\big)^{-1}
\mathbf{y}^{(d)} 
\\
&\quad
-\frac{D}{2}
\log\det\big(K_{XX}(\theta)+\sigma^{2}I\big)
-\frac{ND}{2}\log(2\pi),
\end{aligned}
\end{equation}
where $\theta$ denotes the trainable circuit parameters entering the 
quantum feature map.
This formulation yields a quantum-enhanced GP model framework for predicting multiple outputs, in which the classical covariance function is replaced 
by a quantum similarity measure computed from local reduced-state 
information.

\subsubsection{Projected Kernel Construction}

Following the projected quantum kernel construction introduced in~\cite{b16,b17}, we compare inputs through local reduced density matrices rather than through full-state fidelities. Let
\[
|\psi(x)\rangle = U_{\phi}(x,\theta)\,|0\cdots0\rangle
\]
denote the quantum feature state prepared from input $x$ by the parameterized feature map $U_{\phi}$. Let $\{1,2,\ldots,n_q\}$ denote the full set of qubit indices. Here $n_q$ is the total number of qubits in the feature map circuit, in contrast to the per-stream usage in Section~\ref{sec:kqrc}. For any subset
$\mathcal{K} \subseteq \{1,2,\ldots,n_q\}$ of size $k$, we define the corresponding
$k$-body reduced density matrix as
\begin{equation}
\rho_{\mathcal{K} }(x)
=
\mathrm{Tr}_{\overline{\mathcal{K} }}
\bigl[
|\psi(x)\rangle\langle\psi(x)|
\bigr],
\end{equation}
where $\overline{\mathcal{K} } = \{1,2,\ldots,n_q\}\setminus \mathcal{K} $ is the complement of $\mathcal{K} $. Let $\mathcal{Q}_k(n_q)$ denote the
collection of all $k$-qubit subsets of $\{1,2,\ldots,n_q\}$.
The projected quantum kernel is defined as
\begin{equation}
K_{\mathrm{proj}}^{(k)}(x_i,x_j)
=
\sum_{\mathcal{K} \in \mathcal{Q}_{k}(n_q)}
\mathrm{Tr}\!\left[
\rho_{\mathcal{K} }(x_i)\,\rho_{\mathcal{K} }(x_j)
\right].
\label{eq:proj_kernel}
\end{equation}
Thus, two inputs are considered similar when their encoded quantum states induce similar local reduced states across the selected qubit subsets. For $k=1$, the kernel compares single-qubit reduced states; for $k=2$, it captures pairwise correlations between qubits. In this work, we focus on $k=2$, which provides a good balance between expressive power and computational tractability for correlated time-series data.

\subsubsection{Measurement-Based Kernel Estimation and Circuit Architecture}
\begin{figure}[!tb]
    \centering
    \includegraphics[width=\columnwidth]{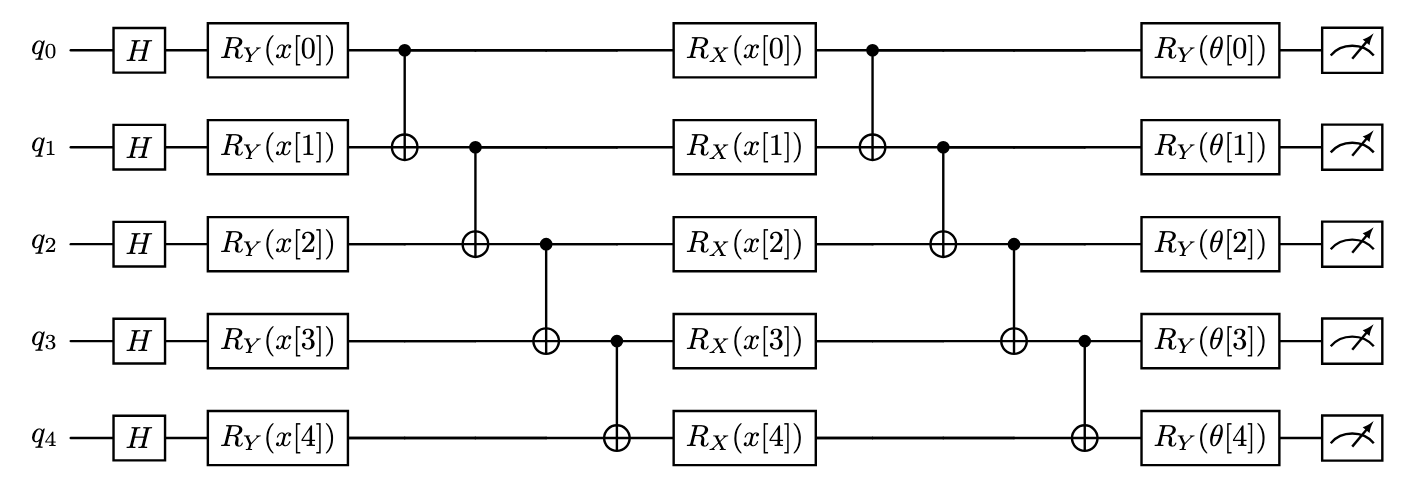}
    \caption{Hardware-efficient feature map used in the QGP model with $n_q=5$ qubits. The circuit alternates data encoding, entanglement, and data re-uploading, producing the state $\lvert \psi(x)\rangle$ used in the projected kernel.} 
    \label{fig:Archi_circuit}
\end{figure}

The projected kernel~\eqref{eq:proj_kernel} can be evaluated from measurement outcomes by expanding each reduced density matrix into the Pauli basis $\mathcal{P}_k = \{I,X,Y,Z\}^{\otimes k}$:
\begin{equation}
\rho_{\mathcal{K}}(x)
=
\frac{1}{2^k}
\sum_{P\in\mathcal{P}_k}
\langle P\rangle_{x,{\mathcal{K}}}\,P,
\end{equation}
where $\langle P\rangle_{x,\mathcal{K} }$ is the expectation value of the Pauli operator $P$ on subset $\mathcal{K} $ for the state encoded by input $x$. Substituting this expansion into Eq.~(\ref{eq:proj_kernel}) gives
\begin{equation}
K_{\mathrm{proj}}^{(k)}(x_i,x_j)
=
\sum_{{\mathcal{K}}\in \mathcal{Q}_k(n_q)}
\frac{1}{2^k}
\sum_{P\in\mathcal{P}_k}
\langle P\rangle_{x_i,{\mathcal{K}}}\,
\langle P\rangle_{x_j,{\mathcal{K}}}.
\end{equation}

These expectation values can be estimated from measurement bitstrings after applying appropriate basis rotations. In practice, this enables direct evaluation of the kernel
using experimentally accessible quantities on near-term quantum hardware.

The projected quantum kernel is defined through a parametrized quantum feature map, which prepares the state $\lvert \psi(x)\rangle$ used in the kernel construction.
We employ a hardware-efficient circuit architecture~\cite{agliardi2025mitigating} designed for near-term quantum devices where the 2-qubit gate ladder is constructed based on shortest distance across connected qubits.
As illustrated in Fig.~\ref{fig:Archi_circuit}, the circuit for $n_q$ $= 5$ qubits consists of a single-layer ansatz composed of alternating data encoding and entangling operations.

Specifically, the circuit begins with a Hadamard initialization layer, followed by a data-encoding layer using $R_Y(x)$ rotations. Entanglement is introduced through a chain of nearest-neighbour CNOT gates. A second encoding layer with $R_X(x)$ rotations
implements data re-uploading~\cite{b22}, followed by a second entangling layer and a
final trainable layer of $R_Y(\theta_i)$ rotations. Measurements are performed on selected qubit subsets to estimate the expectation
values required for kernel evaluation.

The model is trained by minimizing the GP log-likelihood, jointly optimizing the circuit parameters $\theta$ and the noise parameter using the parameter-shift rule~\cite{b23}. Due to the computational
cost of iterative optimization on quantum hardware, all training  is
performed on a quantum simulator. The optimized parameters $\theta^*$  are then
transferred to hardware and kept fixed during the prediction phase.

\subsubsection{Complexity and Practical Reductions}

The number of $k$-qubit subsets grows as $\binom{n_q}{k}$, and for a dataset of $N$
samples the full kernel matrix requires $\mathcal{O}(N^2)$ pairwise evaluations. A native implementation therefore leads to a worst-case complexity of
\begin{equation}
\mathcal{O}\!\left(
N^2\cdot \binom{n_q}{k} \cdot 4^{2k}
\right),
\end{equation}
which becomes prohibitive for large $k$.

In practice, the method remains tractable by restricting to small values of $k$ (typically $k \in \{1,2\}$) and exploiting structure in the measurement process. 
Three implementation strategies reduce the practical
cost well below this bound. First, Pauli operators are grouped into shared measurement bases,
reducing the number of required circuit executions~\cite{b5,b7}. Second,  the set of
qubit subsets is restricted to pairs directly coupled by the circuit entanglement
structure, reducing the number of measured subsystems from $\binom{n_q}{2}$ to
$\mathcal{O}(n_q)$. Third, expectation values are computed once per datapoint and
reused in classical post-processing, shifting most of the computational cost to the
classical domain.

As a result, the effective number of quantum circuit executions scales as 
$
N \times \text{(qubit subsets)} \times \text{(measurement bases)}$,
 which is significantly lower than the naive quadratic cost.

\subsection{Theoretical Conditions for Quantum Performance}
\label{sec:quantum_advantage_theory}

To assess whether the proposed quantum models may  outperform classical counterparts, we adopt the theoretical framework
of Huang~et~al.~\cite{huang2021power}, which characterizes quantum performance through
two complementary diagnostics: the geometric difference between kernel matrices and an effective model-complexity measure.

Given a classical kernel matrix $K_C$ and a quantum kernel matrix $K_Q$, the geometric difference is defined as
\begin{equation}
g_{CQ} = g(K_{C} \parallel K_Q) =
\sqrt{ \left\lVert \sqrt{K_Q} (K_C)^{-1} \sqrt{K_Q} \right\rVert_{\infty} },
\end{equation}
where $\|\cdot\|_\infty$ denotes the spectral norm. This quantity measures how differently the two kernels represent pairwise similarity on the same training set of size $N$.

Small values of $g_{CQ}$ compared with $\sqrt{N}$ indicate that the classical and quantum kernels induce similar geometries, in which case classical models are expected to match or closely approach the quantum model. $g_{CQ} \propto \sqrt{N}$ indicate stronger geometric separation and therefore a greater possibility that the quantum model accesses a function space that is not well captured by the classical kernel.

In our setting, $K_Q$ is the quantum kernel obtained from the KQRC-RM method described in Sec.~\ref{sec:kqrc}, and is constructed via the fidelity-based feature map, while $K_C$ is a classical matrix obtained from standard classical kernels, including RBF, Laplacian, Rational Quadratic, and Mat\'{e}rn
kernels~\cite{scholkopf2002learning, williams2006gaussian}.

The second diagnostic evaluated how simply a given kernel represents the training targets. Following Ref.~\cite{huang2021power}, we quantify the model complexity associated with a kernel matrix $K$ as
\begin{equation}
\kappa(K) = \sum_{i=1}^{N} \sum_{j=1}^{N} \left( K^{-1} \right)_{ij} y_i y_j = \mathbf y^\top K^{-1}\mathbf y,
\end{equation}
where $y_i$ are the target values, and $N$ denotes the number of training samples. Accordingly, the classical and quantum complexities are given by: $\kappa_C = \mathbf y^\top K_C^{-1}\mathbf y$ and $\kappa_Q = \mathbf y^\top K_Q^{-1}\mathbf y$ respectively. A lower value of $\kappa$ indicates that the corresponding kernel induces a simpler effective model for the target mapping. Thus, for a given dataset, a favorable regime for quantum performance is characterized not only by geometric separation between $K_Q$ and $K_C$, but also by $\kappa_Q < \kappa_C$.

In practice, we evaluate $g_{CQ}$, $\kappa_C$, and $\kappa_Q$ as functions of the training-set size $N$. Following the scaling arguments of Ref.~\cite{huang2021power}, evidence for a favorable quantum regime is obtained when $g_{CQ}$ increases with dataset size, the classical complexity scales approximately as $\kappa_C \propto N$, and the quantum complexity remains substantially smaller, ideally $\kappa_Q \ll N$. These diagnostics provide a theoretical pre-screening step before comparing the predictive performance of KQRC-RM and QGP on the Smart Meter forecasting task.

\section{Results}
\label{sec:results}

\subsection{Theoretical Diagnostics of Quantum Performance}
\label{sec:results_advantage}

We first examine whether the Smart Meter dataset satisfies the theoretical conditions associated with improved quantum performance by evaluating the geometric difference $g_{CQ}$ and the model-complexity terms $\kappa_C$ and $\kappa_Q$ defined in Sec.~\ref{sec:quantum_advantage_theory}. For each training set of size $N$, kernel matrices were computed for the quantum representation and the classical baseline kernels.

\begin{figure}[!tb]
    \centering
    \includegraphics[]{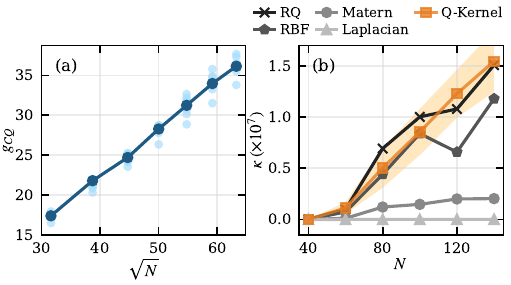}
    \caption{(a) Geometric difference $g_{CQ}$ between the quantum kernel constructed via the fidelity-based feature map and the classical Laplacian kernel as a function of $\sqrt{N}$. Light blue markers denote individual realizations and the dark blue line shows the mean across runs. The approximately linear trend with $\sqrt{N}$ is consistent with increasing geometric separation as the training-set size grows. (b) Effective model complexity $\kappa$ as a function of $N$ for the projected quantum kernel (Q-Kernel) and the classical baselines (Rational Quadratic (RQ), RBF, Mat\'ern, and Laplacian). The shaded band around the Q-Kernel mean indicates $\pm 1$ standard deviation across 10 realizations. Lower values of $\kappa$ correspond to a simpler effective model for the target mapping.}
    \label{fig:g_trend}
\end{figure}

Figure~\ref{fig:g_trend}(a) shows the geometric difference $g_{CQ}$
between the Laplacian classical kernel and the fidelity-based quantum kernel for 10 experimental iterations. Over the explored range of training-set sizes, $g_{CQ}$ increases approximately linearly with $\sqrt{N}$, indicating progressively stronger geometric separation between the classical and quantum representations.
Within the framework of~\cite{huang2021power}, this behavior is consistent with a regime in which the quantum model accesses a function space that is increasingly distinct from that of the classical baseline, and therefore with conditions under which improved quantum performance may become plausible. 

The classical and quantum model complexities, $\kappa_C$ and $\kappa_Q$, as functions
of dataset size $N$, are presented in  Figs.~\ref{fig:g_trend}(b). 
 The classical complexity exhibits a marked dependence on the choice of kernel. In particular, the RBF and Rational Quadratic kernels show a strong increase with $N$, the Mat\'{e}rn kernel increases more moderately, and the Laplacian kernel remains substantially smaller over the explored range. The quantum complexity $\kappa_Q$ also increases with $N$, with an approximately linear trend in the mean across realizations. Relative to the classical baselines, $\kappa_Q$ is lower than the RBF and Rational Quadratic complexities at larger dataset sizes, but remains higher than the Laplacian and Mat\'{e}rn cases. Taken together, these results suggest that the Smart Meter dataset exhibits partial signatures consistent with the theoretical conditions for favorable quantum performance, although the evidence is not uniform across all classical kernel baselines. The present results should therefore be interpreted as indicative rather than conclusive, and as motivation for the empirical performance analysis of KQRC-RM and QGP in the following subsections.

\subsection{KQRC-RM Results}
\label{sec:results_kqrc}

\begin{figure}[!b]
    \centering
    \includegraphics[]{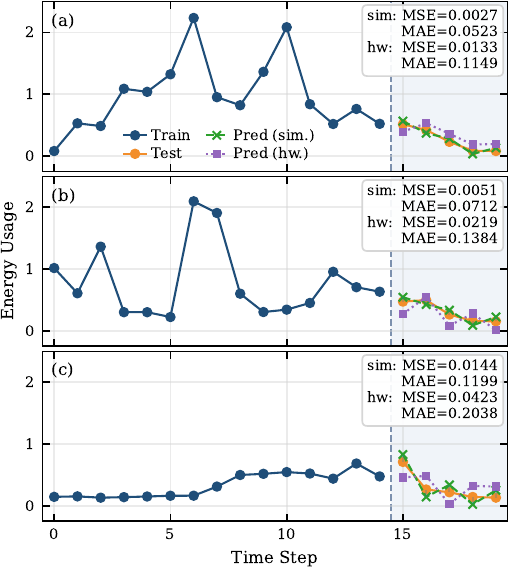}
    \caption{KQRC-RM experiments: Forecasting results for the representative triplet: (a) Customer 4, (b) Customer 8, and (c) Customer 11, comparing predictions obtained with the MPS simulator and  real quantum hardware. The dashed vertical line marks the end of the training window,
    and the shaded region indicates the 5-step forecast horizon.}
    \label{fig:sim_hardware_results}
\end{figure}

\begin{figure*}[tb]
    \centering
    \includegraphics[]{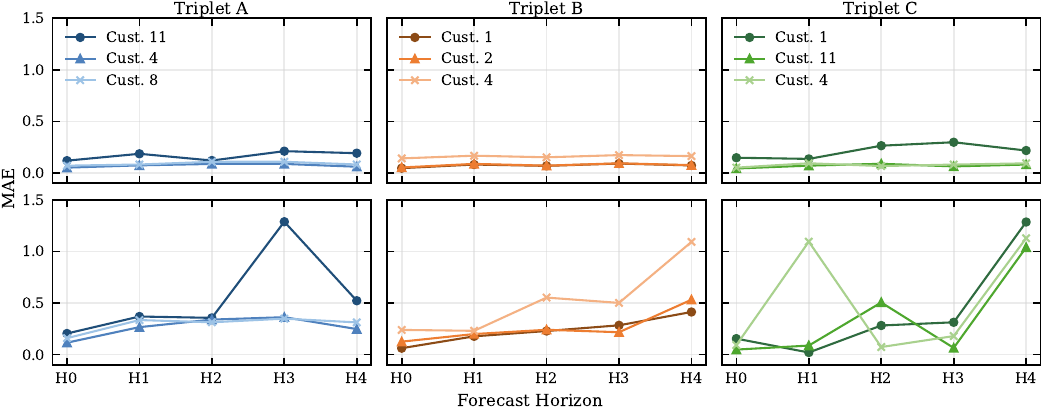}
    \caption{KQRC-RM experiments: Per-horizon MAE across a five-hour forecast horizon for the three customer triplets, arranged by columns: Triplet A (Customers 4, 8, and 11), Triplet B
    (Customers 1, 2, and 4), and Triplet C (Customers 11, 4, and 1). The top row show results obtained with the MPS simulator, and the bottom row shows the corresponding results on real quantum hardware.}
    \label{fig:batch_combined}
\end{figure*}

We evaluate the KQRC-RM framework on real quantum hardware using three-customer
subsets processed in parallel. Each customer stream is mapped to an
independent system register, while cross-stream dependencies are captured through
chained inter-stream entangling operations, as shown in Fig.~\ref{fig:arch}. Guided by the
empirical qubit-scaling study in Appendix Fig.~\ref{fig:ent}, we fix the
reservoir size to 15 system qubits per customer. This choice places the model well within the regime
where reservoirs with cross-stream entanglement already achieve consistently low
average relative error (below 0.30 from approximately 8 qubits onward), after which
performance saturates only gradually.  This corresponds to 30 qubits per customer stream (15 system and 15 ancilla), or a total of 
90 qubits for three concurrent streams.

\paragraph{Hardware vs.\ simulator performance}
For the KQRC-RM experiments, we prepared three customer triplets selected according to correlation-based subset construction described in Appendix~\ref{app:dataset}. Forecasting experiments were performed on both a Matrix Product State (MPS) simulator and the
\texttt{ibm\_marrakesh} superconducting quantum processor (Appendix~\ref{app:hardware}), using the same circuit
architecture, kernelization procedure, and train-test split for each triplet. Models were trained
on a rolling window of 15 hours of historical data and evaluated by forecasting the
subsequent 5 hours. On hardware, we enabled dynamical decoupling using the \texttt{XpXm}
pulse sequence and gate twirling with 32 randomizations. The transpiled circuit had a total depth of $114$, 
a two-qubit gate depth of 51, and gate counts; 471 SX, 315 CZ, 195 RZ, and 93 RX.

Figure~\ref{fig:sim_hardware_results} shows a representative example for triplet comprising Customers 4, 8, and 11. The simulator predictions closely follow the target trajectories across all three customers, indicating that the reservoir and kernelized readout capture the dominant dynamics of the selected streams. This visual trend is consistent with the error values reported in the figure, where the simulator achieves lower MSE and MAE than hardware for all three customers.
The hardware results exhibit larger deviations,
primarily due to gate noise, readout error, decoherence, and accumulated error from
repeated entanglement and measurement cycles. Despite this degradation, the
hardware forecasts still capture the main temporal structure and directional
trends of the energy-consumption patterns, indicating that the repeated-measurement reservoir dynamics remain operational on real quantum hardware.

\paragraph{Triplet-wise horizon analysis}
Figure~\ref{fig:batch_combined} presents the per-horizon MAE
for the three customer triplets, enabling a direct comparison of multi-step forecasting
behavior across forecast horizons. 
In the MPS simulator, the error profiles remain comparatively low and stable across most horizons, although individual
customers show localized increases in MAE in some triplets.
On hardware, the same
overall trends are accompanied by substantially larger horizon-dependent fluctuations.Triplet A exhibits
consistently low and smooth MAE under the MPS simulator across all horizons, while the
hardware results show a moderate increase in error that becomes more pronounced at
longer horizons. Triplet B demonstrates higher overall MAE for both platforms,
indicating a more challenging forecasting setting. Triplet C represents the most
difficult case, where the divergence between simulator and hardware performance is
largest and the hardware MAE shows greater variability at extended horizons.
The per-customer variations across triplets reflect the underlying correlations between the co-encoded
series, since the three customers in each triplet are entangled together through CNOT gates before
the ancilla measurements. Different triplets therefore expose each customer to different correlation
patterns among its partners, which shape the effective feature representation seen by the readout.
Nevertheless, the
qualitative trend structure remains consistent between the two environments, suggesting
that the underlying recurrent reservoir dynamics remain effective even under realistic
noise conditions.

\subsection{QGP Results}
\label{sec:results_qgp}

Since the qubit requirement of QGP with a projected quantum kernel scales linearly with the number of customer time-series, using one qubit per customer, it is substantially more hardware-efficient than KQRC-RM. This enables us to move beyond
the three-customer subsets considered for KQRC-RM and evaluate QGP in larger
multi-customer forecasting settings. Accordingly, we focus on 5-customer subsets and a utility-scale experiment
involving 100 customers. All experiments are evaluated on both a quantum simulator and real quantum
hardware (\texttt{ibm\_marrakesh}).

\paragraph{Hardware vs. simulator performance.}
For the 5-customer configuration, QGP is used in a multi-input, multi-output
forecasting setting, where each training sample is a 5-dimensional vector of
simultaneous hourly demand values from five customers. For each group, the training
set consists of 15 consecutive multivariate hourly observations, and the
model is used to predict the subsequent 5 multivariate hourly observations. We
evaluate three fixed 5-customer groups, as described in Appendix~\ref{app:dataset}, and report results averaged
over five realizations for each group.

\begin{figure}[!t]
    \centering
    \includegraphics[]{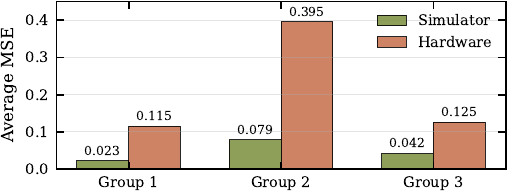}
    \caption{QGP experiments: Five-hour-ahead multi-output prediction performance analysis across three 5-customer groups on both simulator and quantum hardware. Bars report the average MSE over the 5-hour forecast interval. Results for each group are averaged over five realizations.}    \label{fig:prediction_analysis_1}
\end{figure}

The prediction results for both simulator and hardware are summarized in Fig.~\ref{fig:prediction_analysis_1}. For a quantum simulator results, the model achieves consistently low
MSE across all tested customer groups, indicating stable predictive performance. On quantum hardware, the same overall trend is preserved, although
the error becomes more sensitive to the selected customer subset and generally
increases relative to simulation. This behaviour is consistent with the effect of
noise and decoherence during kernel estimation and circuit execution. Despite this
degradation, the hardware results remain reasonably close to the simulator, indicating that the projected-kernel model retains useful predictive
capability under realistic device noise. Representative QGP forecasts, together with their predictive uncertainty, are shown in Appendix Fig.~\ref{fig:gaussian_process_vertical}(b)-(g). Loss convergence is shown in Appendix Fig.~\ref{fig:gaussian_process_vertical}(a). 



\begin{table}[!bt]
\centering
\caption{QGP utility-scale experiment: Performance for the
100-qubit hardware run. Customers are partitioned into accuracy tiers according to
their MAE and MSE values.}
\label{tab:100qubits_performance}
\begin{tabular}{lcccc}
\hline
\textbf{Tier} & \textbf{MAE Range} & \textbf{Share} & \textbf{Avg MAE} & \textbf{Avg MSE} \\
\hline
Low    & $<0.15$        & 49\% & 0.082 & 0.011 \\
Medium & $0.15$--$0.35$ & 31\% & 0.229 & 0.100 \\
High   & $>0.35$        & 20\% & 0.664 & 0.909 \\
\hline
\end{tabular}
\end{table}

\paragraph{Utility-scale 100-qubit experiment.}
To examine the scalability of QGP on real hardware, we performed a utility-scale experiment using data from 100 customers, corresponding to a 100-qubit circuit  with one qubit assigned to each customer time-series. We used a fixed train-test split, with 15 hours of data used for training and 5 hours used for testing.
The hardware execution made use of dynamical decoupling with the \texttt{XpXm}
pulse sequence and gate twirling with 32 randomizations for error supression and migitation. The transpiled circuit had a total depth of 19 with a two-qubit gate depth of 4, and overall gate counts; 625 SX, 533 RZ, 198 CZ, and 18 X.

The resulting prediction errors show a structured distribution across three accuracy tiers, summarized in Table~\ref{tab:100qubits_performance}. Specifically, 49\% of customers fall into a low-error regime, 31\% into a medium-error regime, and 20\% into a high-error regime. Thus, in this single large-scale run, 80\% of customers lie in
the low- or medium-error groups. 

To understand the spatial origin of these accuracy tiers, Figure~\ref{fig:ibm_marrakesh} shows the topology of the \texttt{ibm\_marrakesh} processor, where the node color encodes the T2 dephasing time of each physical qubit (green: long coherence, red: short coherence) and the edge color encodes the CZ gate error rate of each two-qubit link (green: low error, red: high error). The 100 qubits used in this experiment are mapped contiguously across the processor, so each customer's prediction quality is directly determined by the hardware quality of the qubit assigned to it.
 Customers assigned to qubits 3 to 15, characterized by $T_2 > 150\,\mu\text{s}$ and CZ error rates below 0.2\% tend to fall into the low-error tier, whereas customers assigned to qubits in degraded regions, particularly the middle chain (physical qubits 41 to 55) and the bottom strip (physical qubits 121 to 155), where $T_2$ values fall as low as 14 to 24 $\mu\text{s}$ and CZ error rates reach up to 1.27\% account for the majority of the high-error. Taken together, the accuracy-tier distribution is therefore not random: it reflects the heterogeneous noise landscape of the physical device, and the 20\% high-error rate is largely attributable to hardware limitations rather than to fundamental limitations of the quantum Gaussian process model itself.

This outcome demonstrates that the utility-scale experiment on 100 qubits is sufficiently robust to produce useful predictions for the vast majority of customers, supporting the feasibility of quantum-assisted forecasting in practical settings. Even under the noise and connectivity constraints inherent to current quantum hardware, the model successfully captures demand patterns for 4 out of every 5 customers, providing a strong indication that quantum Gaussian process regression at the 100-qubit scale is a viable approach for real-world forecasting tasks. The remaining 20\% of customers in the high-error regime are attributable primarily to hardware noise specifically, short T2 coherence times and elevated CZ gate error rates on the physical qubits to which they were assigned though non-stationary, or highly irregular demand patterns may compound this effect for a subset of cases. Improved qubit selection strategies or hardware-aware transpilation targeting the highest-quality regions of the processor would be expected to reduce this fraction substantially.

\section{Benchmarking Against Classical Models}
\label{sec:benchmarking}

\begin{figure}[!bt]
\centering
\includegraphics[]{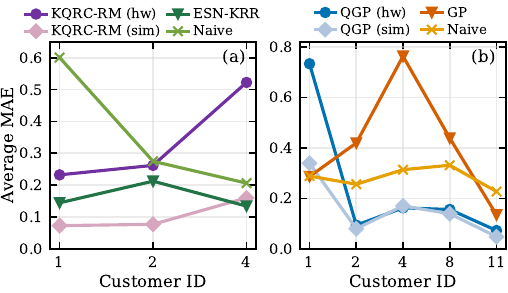}
\caption{ Benchmarking comparison of quantum and classical forecasting models in small-scale setting.
(a) KQRC-RM performance on hardware and simulator compared with ESN-KRR and the
na\"{i}ve persistence baseline, showing average MAE per customer over five
realizations. (b) QGP performance on hardware and simulator compared with classical
multi-output GP and the na\"{i}ve baseline, showing average MAE per customer over
five 5-hour test subsets.}
\label{fig:benchmarking_plot}
\end{figure}

We benchmark the KQRC-RM and QGP models for energy consumption forecasting against
classical approaches. Within each experimental setting, the quantum and classical models use the identical training and testing splits. 
The benchmarking results
reported in this section are focus on small-scale configurations: 
3-customer subset for KQRC-RM and 5-customer subset for QGP. 

For KQRC-RM, we use an Echo State Network with kernel ridge regression (ESN-KRR) as
the classical baseline~\cite{2001Jaeger,2021Nakajima,2024Sun,1998Saunders,
2002Scholkopf}. This provides a standard classical analogue to reservoir computing
while keeping the readout mechanism comparable. Unlike the quantum model, which
processes customers jointly, the ESN baseline trains a separate reservoir for each
customer. The benchmark is performed on Triplet B (Customers 1, 2, and 4) using a
rolling-window dataset with 15 hours of training data followed by a 5-hour forecast
horizon per customer. The evaluation is repeated over five realizations
corresponding to different train--test windows, and the results shown in
Fig.~\ref{fig:benchmarking_plot}(a) give the average MAE per customer across these
realizations. We additionally include the na\"{i}ve persistence baseline, which uses
the last observed value as the forecast for all future time steps. As shown in
Fig.~\ref{fig:benchmarking_plot}(a), for selected Triplet B dataset, the simulator yields the lowest errors overall,
while the hardware implementation remains competitive with ESN-KRR for some
customers but shows larger degradation for others. Averaged over the three customers in Triplet B, KQRC-RM achieves an MAE of 0.10 on the simulator and 0.34 on hardware, compared with 0.16 for ESN-KRR and 0.36 for the na\"{i}ve persistence baseline. Thus, the simulator reduces
the average MAE by 36.92\% relative to ESN-KRR, while the hardware implementation has a 107.08\% higher average MAE than ESN-KRR, but remains slightly below the na\"{i}ve baseline.

For QGP, we use a classical Gaussian Process baseline with an RBF kernel in a
multi-output formulation based on a linear model of
coregionalization~\cite{2006Rasmussen,2007Bonilla,2012Alvarez}. This baseline is
intended to match the multi-customer prediction setting of QGP more closely by
allowing correlations across customer outputs. The comparison is carried out for
Group A (Customers 1, 2, 4, 8, and 11) using 15 consecutive hourly observations per
customer as input and the subsequent 5 hours as the prediction horizon. The
evaluation is repeated over five different test subsets, and
Fig.~\ref{fig:benchmarking_plot}(b) shows the average MAE per customer across these
realizations. We also include the na\"{i}ve persistence baseline as a
reference. As shown in Fig.~\ref{fig:benchmarking_plot}(b), for the selected Group A dataset, both simulator and
hardware QGP remain competitive with the classical multi-output GP across several
customers, although the relative performance varies across the selected group. Averaged over the five customers in Group A, QGP achieves an MAE of 0.16 on the simulator and 0.24 on hardware, compared with 0.41 for the classical multi-output GP and 0.28 for the na\"{i}ve persistence baseline. This corresponds to average MAE reductions of 62.01\% on the simulator and 40.37\% on hardware
relative to the classical GP baseline.

\section{Conclusion}
\label{sec:conclusion}

This work introduced two hybrid quantum-classical frameworks for multi-output
time-series forecasting in energy systems: Kernelized Quantum Reservoir Computing
with Repeated Measurement (KQRC-RM) and a Projected Quantum Kernel Gaussian Process
(QGP). Both approaches were evaluated on real quantum hardware and compared with
classical baselines on a Smart Meter dataset of 103 customer electricity-consumption
time-series.

The results show that both models remain operational under realistic NISQ
constraints, while exhibiting the expected performance degradation on hardware
relative to simulation. KQRC-RM demonstrates that repeated measurement, coupled
quantum reservoirs, and kernelized readouts can be integrated into a single
multi-stream forecasting framework capable of capturing both temporal structure and
cross-stream correlations. At the same time, its resource requirements are
substantial, reaching 114 qubits for the three-stream setting considered here. In the small-scale benchmarking study, KQRC-RM reduces the average MAE relative to the ESN-KRR baseline by 36.92\% on the simulator, although its hardware implementation remains 107.08\% higher in MAE, reflecting the sensitivity of the present circuit implementation to device noise.

The QGP framework provides a more hardware-efficient and scalable alternative.
By replacing fidelity-based kernels with projected quantum kernels constructed from
local reduced-state statistics, it enables multi-output Bayesian forecasting with
predictive uncertainty while improving robustness to hardware noise. 
Across the small-scale benchmarking experiments, QGP reduces the average MAE relative to the classical multi-output GP baseline by 62.01\% on the simulator and 40.37\% on hardware. In the utility-scale 100-qubit experiment, 49\% of customers fall into a
low-error regime and 80\% into low- or medium-error regimes overall. Notably, we used a topology-aware circuit optimization which gives rise to a shallow ansatz at utility scale, yet also providing an overall accuracy improvement compared to the classical baselines.


Taken together, these results show that hybrid quantum forecasting can already be
studied experimentally on practical time-series tasks at meaningful scales.
However, the present findings should be interpreted primarily as an empirical
feasibility study rather than as evidence of practical advantage over
classical forecasting methods. Comparing KQRC-RM and the QGP models, KQRC-RM should be used when the training dataset is large, yet the number of time-series to predict is small. Whereas, QGP is potentially the better choice when there is a small amount of training data and a large number of time-series to simultaneously predict.


\section*{Acknowledgments}

This work was supported through the partnership between E.ON Group Innovation GmbH
and The Washington Institute for STEM Entrepreneurship and Research (WISER), which
provided the fellowship enabling this research. AL and CO thank Kavitha Yogaraj for
useful insights into the classical and quantum model complexity analysis. The authors thank Timothee Dao for help with the utility scale plot creation and Ege Yilmaz for useful comments to improve the article.

\bibliographystyle{apsrev4-2}
\bibliography{ref}

\newpage
\onecolumngrid
\appendix

\section{Hardware Platform}
\label{app:hardware}

All quantum hardware experiments reported in this work were executed on
\texttt{ibm\_marrakesh} (Heron r2), a 156-qubit superconducting quantum processor
based on transmon qubits. At the time of execution, the device exhibited median
coherence times of $T_1 = 192.39~\mu\text{s}$ and $T_2 = 96.89~\mu\text{s}$, a
median controlled-Z (CZ) gate error rate of $2.334\times10^{-3}$, and a median
readout error of $1.23\times10^{-3}$. The qubits are arranged in a heavy-hexagon
connectivity topology forming a degree-3 lattice, which provides relatively high
connectivity while remaining compatible with superconducting-device fabrication
constraints~\cite{mckay2017}.

\begin{figure}[h]
\centering
\includegraphics[width=0.4\linewidth]{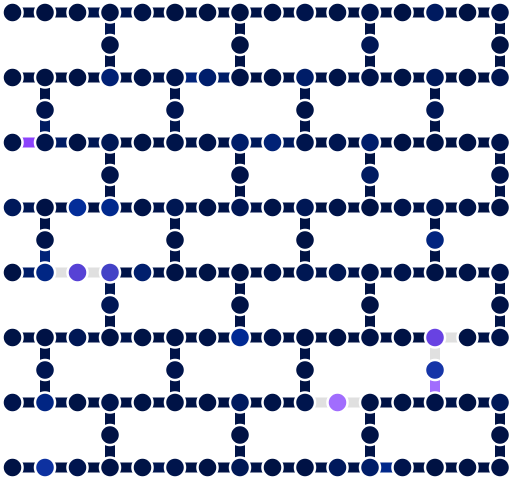}
\caption{Layout of the \texttt{ibm\_marrakesh} processor with calibration data used
for qubit selection and circuit mapping. Lighter qubit colors indicate higher readout
assignment error, while lighter edge colors indicate higher controlled-Z gate error.}
\label{fig:layout}
\end{figure}

Figure~\ref{fig:layout} shows the processor layout together with per-qubit readout
error and per-edge CZ gate error calibration data. These calibration data were used
to guide qubit selection and circuit mapping in the hardware experiments in order to
reduce the impact of device imperfections.

\section{Dataset and Subset Construction}
\label{app:dataset}
This appendix provides supplementary information on the Smart Meter dataset and the
correlation-based construction of customer subsets used in the QGP and KQRC-RM
experiments.

\begin{figure}[h]
    \centering
    \includegraphics[]{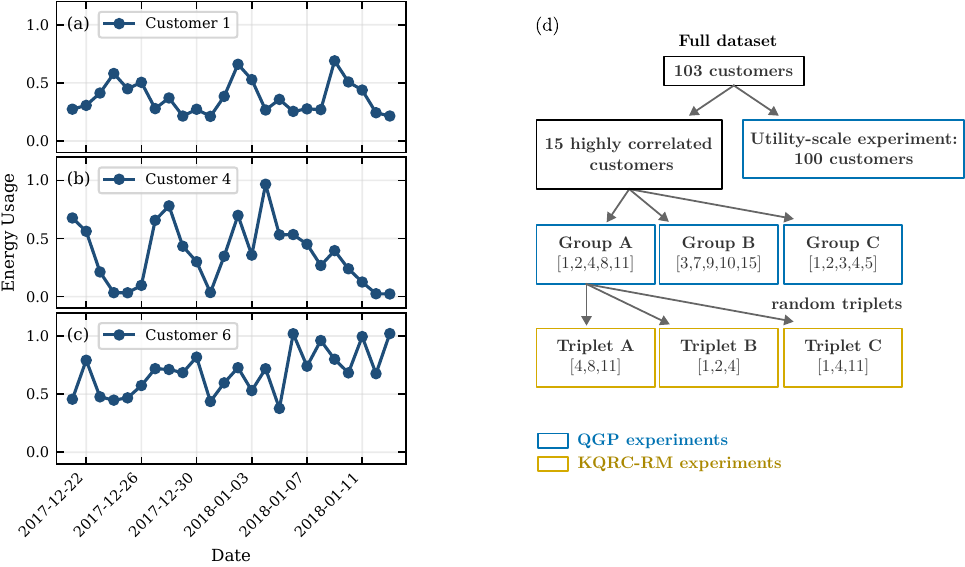}
    \caption{(a)-(c) Representative daily energy-usage profiles for three selected customers from the Smart Meter dataset. The series are obtained by averaging the hourly demand values within each day. (d) Correlation-based construction of customer subsets for the QGP (blue frames) and KQRC-RM (yellow frames) experiments. Starting from the full dataset of 103 customers, a 15-customer correlated subset is selected and then partitioned into three 5-customer groups for QGP. Three 3-customer triplets are further constructed from Group A for the KQRC-RM experiments. A separate 100-customer configuration is used for the utility-scale QGP experiment. 
    The customer groups and triplets are not mutually exclusive partitions of the dataset. Instead, they are independently selected experimental subsets used for different QGP and KQRC-RM benchmarking configurations. Therefore, the same customer may appear in multiple groups or triplets.}
    \label{fig:splitting_customers}
\end{figure}

The dataset consists of hourly electricity-demand measurements collected during the
winter season for 103 residential customers. Figure~\ref{fig:splitting_customers}(a)-(c) shows a
representative sample of four time-series from the selected 15-customer subset,
illustrating the multi-scale variability, local fluctuations, and intermittent peaks
characteristic of residential smart meter data.

To enable structured multi-output forecasting, we construct customer subsets based on
pairwise correlation. The full workflow is summarized in
Fig.~\ref{fig:splitting_customers}(d). Starting from the full dataset of 103 customers,
we first select a reduced set of 15 customers with the strongest pairwise
correlations. This 15-customer subset is used as the basis for the QGP experiments
and for constructing smaller subsets for KQRC-RM.

For the QGP experiments, we consider three fixed 5-customer groups derived from the
15-customer subset: Group A $\{1,2,4,8,11\}$, Group B $\{3,7,9,10,15\}$, and Group C
$\{1,2,3,4,5\}$. In addition, a utility-scale QGP experiment is performed on a
100-customer configuration drawn directly from the full dataset.
For the KQRC-RM experiments, smaller subsets are used because the repeated-measurement
reservoir architecture requires substantially greater hardware resources. We therefore
construct three customer triplets from Group A, namely Triplet A $\{4,8,11\}$,
Triplet B $\{1,2,4\}$, and Triplet C $\{1,4,11\}$, as shown in
Fig.~\ref{fig:splitting_customers}(d). These triplets are used in the multi-stream reservoir forecasting experiments reported in the main text.

\newpage
\section{Supporting Results for KQRC-RM}
\label{app:kqrc}

\begin{figure}[h]
    \centering
    \includegraphics[width=0.4\linewidth]{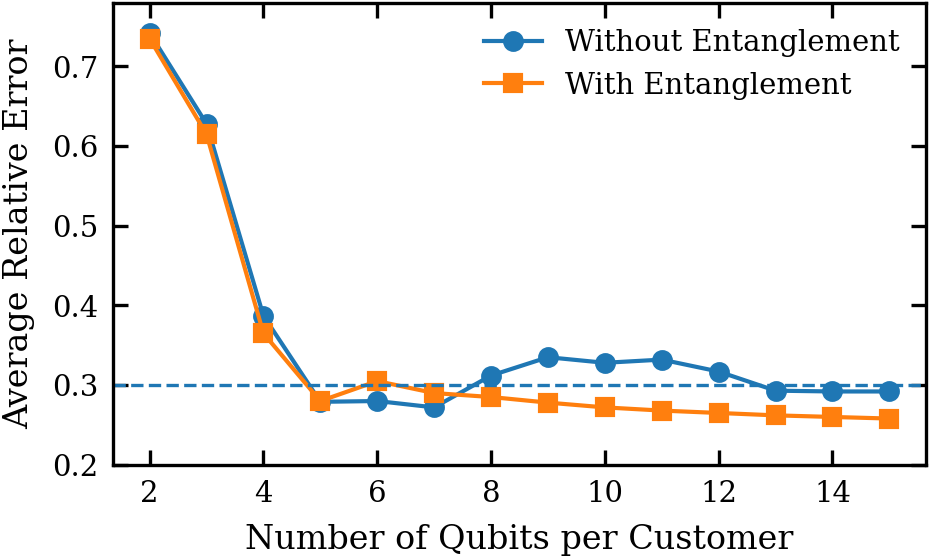}
    \caption{Average relative forecasting error as a function of the number of system
qubits per customer for KQRC-RM reservoir architectures with and without cross-stream
entanglement. The dashed horizontal line marks the relative-error threshold of 0.30.}
    \label{fig:ent}
\end{figure}

This appendix presents the qubit-scaling analysis used to select the reservoir size
for the KQRC-RM experiments. Figure~\ref{fig:ent} shows the average relative
forecasting error as a function of the number of system qubits per customer for
reservoir architectures with and without cross-stream entanglement.
For both architectures, the error decreases sharply as the number of qubits increases
from 2 to about 5, indicating that larger reservoirs substantially improve the
quality of the extracted temporal features in the small-qubit regime. Beyond this
point, the architecture with cross-stream entanglement achieves consistently lower
error than the non-entangled variant. In particular, from approximately 8 qubits
onward, the entangled reservoir remains below the 0.30 relative-error threshold and
shows a stable low-error trend, whereas the non-entangled architecture exhibits
larger fluctuations and generally higher error over the same range.
Based on this behavior, we fix the KQRC-RM reservoir size to 15 system qubits per
customer in the main experiments. This choice places the model within the stable
low-error regime of the entangled architecture while remaining close to its best
observed performance.

\section{Supporting Results for QGP}
\label{app:qgp}
This appendix collects supplementary figures for the QGP experiments described in
Sec.~\ref{sec:results_qgp}. The material is organized in the order in which it is
referenced in the main text: first, the training-loss convergence of the QGP model,
and second, representative prediction results with predictive uncertainty for selected
customers in the small-scale and utility-scale settings.

\begin{figure*}[h]
    \centering
    \includegraphics[]{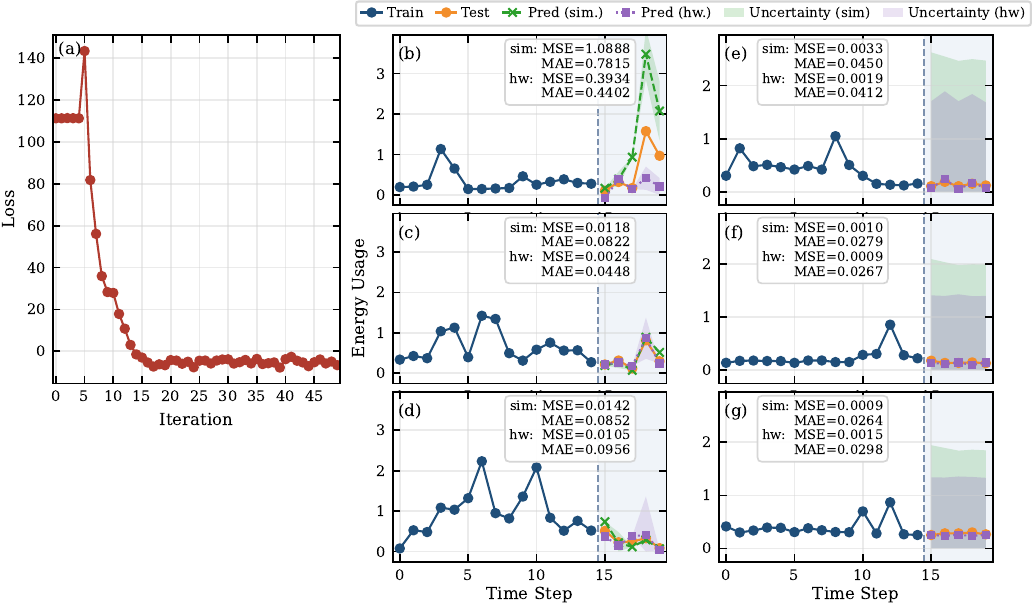}
    \caption{QGP experiments: Representative forecasting results for selected customers in the small-scale and utility-scale settings. (a) Convergence of the negative log-likelihood training loss, optimized with respect to the trainable feature-map parameters $\theta$. (b)–(d) Forecasting results for three customers from the 5-customer Group A experiment. (e)–(g) Forecasting results for three randomly selected customers from the 100-customer utility-scale experiment. In each forecasting panel, the blue curve denotes the training data, the orange curve denotes the test data, the green markers denote simulator predictions, and the purple markers denote hardware predictions. The light-green and light-purple shaded regions indicate the predictive uncertainty for the simulator and hardware results, respectively. The dashed vertical line marks the boundary between the training window and the forecast horizon. Insets report the corresponding MSE and MAE values for the simulator and hardware predictions.}
    \label{fig:gaussian_process_vertical}
\end{figure*}

\paragraph{Training loss convergence.}
The QGP model is trained on a quantum simulator by minimizing the negative
log-likelihood of the Gaussian Process predictive distribution using the
parameter-shift rule. The resulting convergence curve is shown in
Fig.~\ref{fig:gaussian_process_vertical}(a). After an initial transient, the loss decreases rapidly and
then stabilizes, indicating successful optimization of the trainable circuit
parameters in the 5-qubit configuration.

\paragraph{Representative predictions and predictive uncertainty.}
In addition to point forecasts, the QGP framework provides predictive uncertainty
through the Gaussian Process posterior. Figure~\ref{fig:gaussian_process_vertical}(b)-(g)
shows representative forecasting results for selected customers in both the
5-customer and 100-customer experiments. The plotted uncertainty bands widen beyond
the training window, reflecting the increasing uncertainty over the forecast horizon.
The figure also enables direct comparison between simulator and hardware predictions,
showing that the overall forecast structure is largely preserved on real hardware,
although with larger deviations in some cases.

\begin{figure*}[h]
    \centering
    \includegraphics[width=0.8\linewidth]{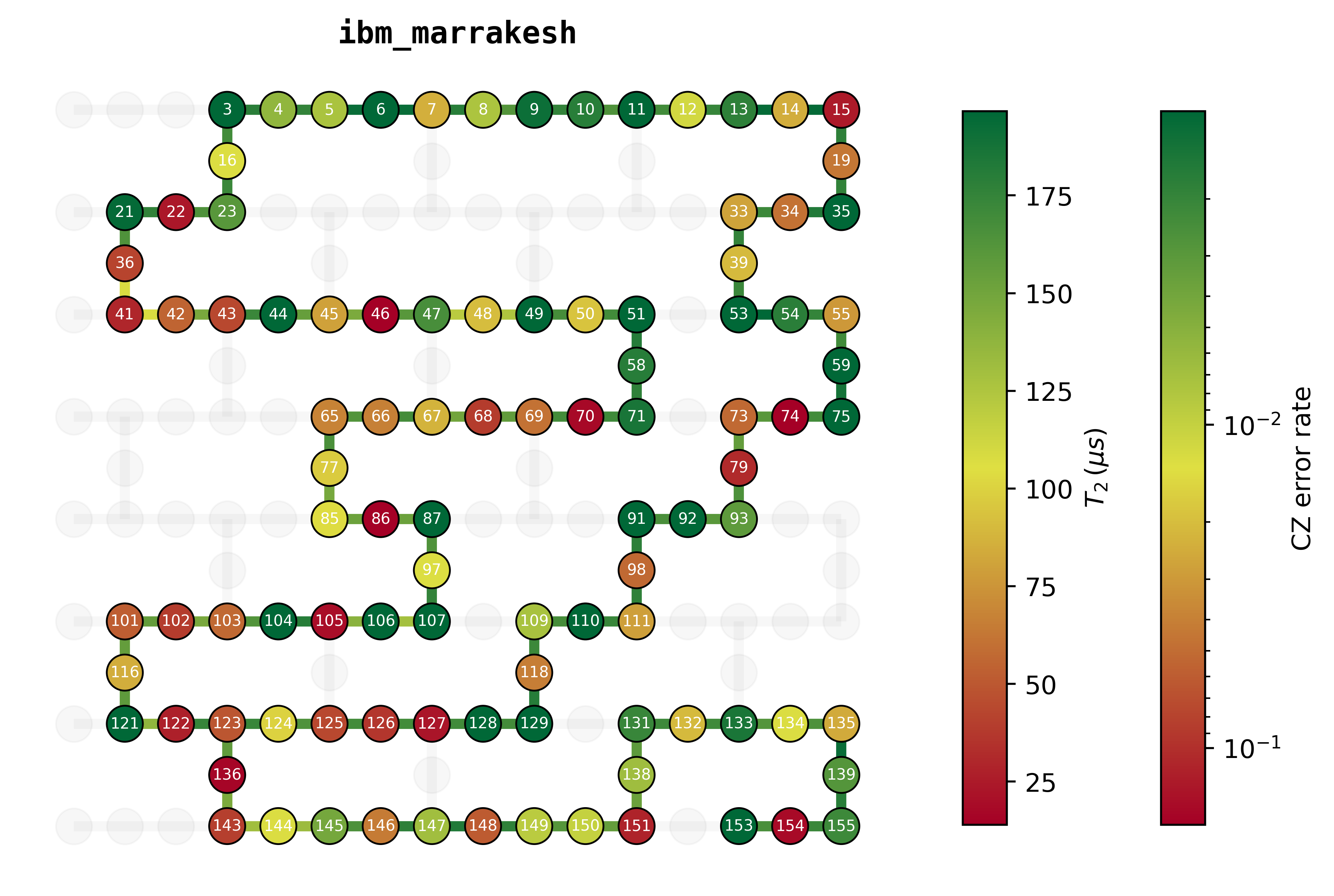} 
    \caption{Hardware noise landscape of the IBM Marrakesh processor for the 100 qubits experiment for the QGP model. Node colour encodes \(T_2\) dephasing time and edge colour encodes CZ gate error rate (green: low noise; red: high noise). The spatial distribution of hardware quality is directly reflected in the prediction errors of Table~\ref{tab:100qubits_performance}.}
    \label{fig:ibm_marrakesh}
\end{figure*}

\begin{figure*}[h]
    \centering
    \includegraphics[width=0.8\linewidth]{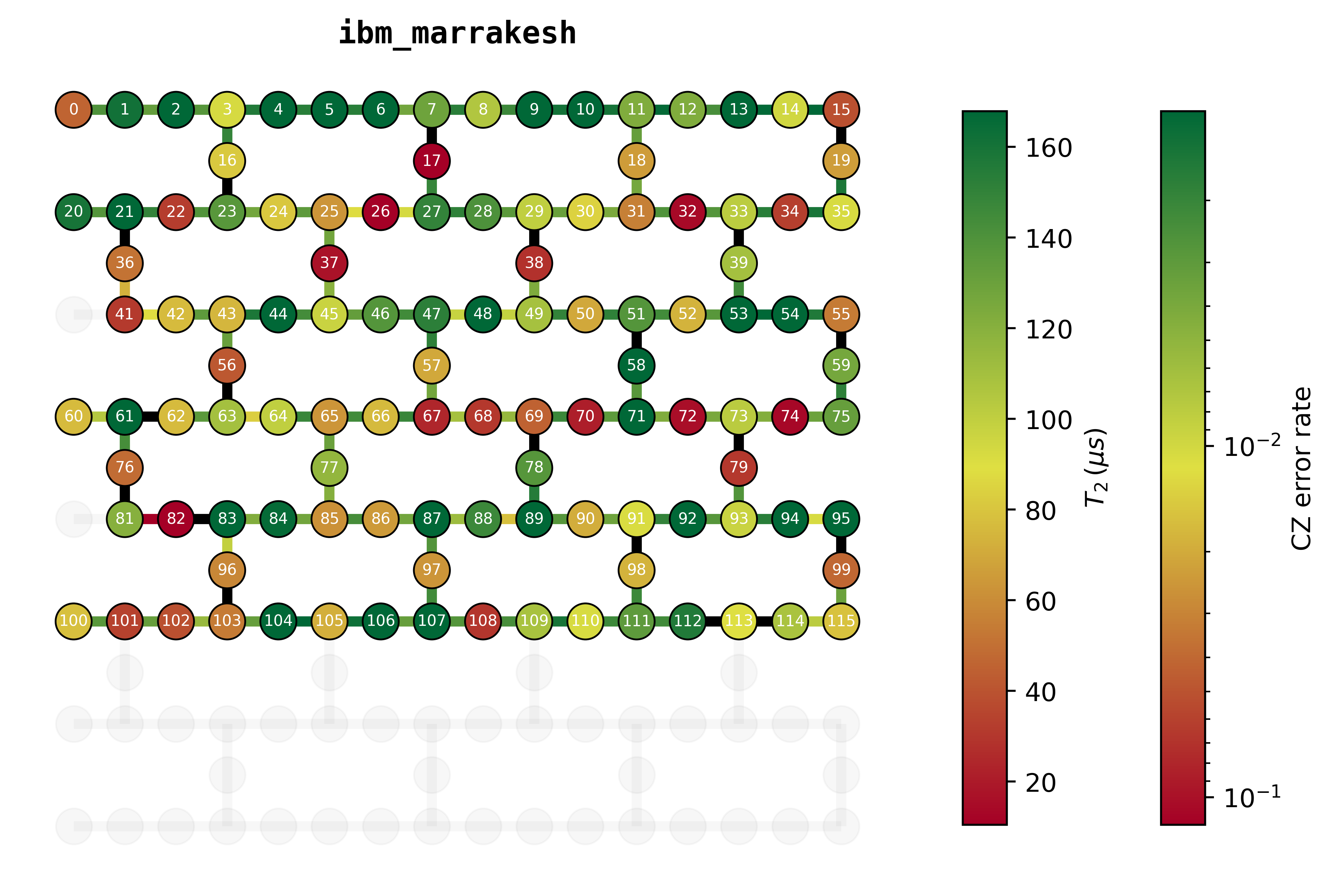} 
    \caption{Hardware noise landscape of the IBM Marrakesh processor for the 114 qubits experiment for the QRC model. Node colour encodes \(T_2\) dephasing time and edge colour encodes CZ gate error rate (green: low noise; red: high noise).}
    \label{fig:ibm_marrakesh1}
\end{figure*}



\end{document}